\documentclass{pnastwo}
\usepackage{color}
\usepackage{float, caption}

\newcommand*\diff{\mathop{}\!\mathrm{d}}
\url{http://www.pnas.org/cgi/doi/10.1073/pnas.1503741112}
\copyrightyear{2015}
\issuedate{Issue Date}
\volume{Volume}
\issuenumber{Issue Number}
	
\begin{document}

\title{Interatomic repulsion softness directly controls the fragility of supercooled metallic melts}

\author{
Johannes Krausser
\affil{1}{Physics Department and Institute for Advanced Study, Technische Universit\"{a}t M\"{u}nchen, 85748 Garching, Germany},
Konrad Samwer
\affil{2}{I. Physikalisches Institut, Universit\"at G\"ottingen, Friedrich-Hund-Platz 1, 37077 G\"ottingen, Germany},
Alessio Zaccone
\affil{1}{}
}

\contributor{Submitted to Proceedings of the National Academy of Sciences
of the United States of America}

\significancetext{Bulk metallic glasses are the most promising materials in many technological applications thanks to their mechanical properties. The stability and thermo-elasticity of supercooled liquid metals is encoded in the temperature-dependence of the viscosity at the glass transition: the fragility. While with colloidal glasses it has been possible to explain the fragility in terms of the "softness" (or its inverse: the steepness) of the microscopic inter-particle potential, the same could not be done with metals due to the complex interatomic interaction. Here we solve this problem and propose a new methodology which provides the missing analytical link between fragility and interatomic potential in metals. Surprisingly, our results show that the same scenario found earlier with colloidal glasses applies to metals too.}

\maketitle

\begin{article}

\begin{abstract}
We present an analytic scheme to connect the fragility and viscoelasticity of metallic glasses to the effective ion-ion interaction in the metal.
This is achieved by an approximation of the short-range repulsive part of the interaction, combined with nonaffine lattice dynamics to obtain analytical expressions for the shear modulus, viscosity, and fragility in terms of the ion-ion interaction.
By fitting the theoretical model to experimental data, we are able to link the steepness of the interionic repulsion to the Thomas-Fermi screened Coulomb repulsion and to the Born-Mayer valence-electron overlap repulsion for various alloys.
The result is a simple closed-form expression for the fragility of the supercooled liquid metal in terms of few crucial atomic-scale interaction and anharmonicity parameters.
In particular, a linear relationship is found between the fragility and the energy scales of both the screened Coulomb and the electron-overlap repulsions.
This relationship opens up opportunities to fabricate alloys with tailored thermo-elasticity and fragility by rationally tuning the chemical composition of the alloy according to general principles.
The analysis presented here brings a new way of looking at the link between the outer-shell electronic structure of metals and metalloids and the viscoelasticity and fragility thereof.

\end{abstract}
\keywords{metallic glasses | fragility | repulsive interaction}
\section{Introduction}
\dropcap{U}nderstanding the mechanism which governs the emergence of mechanical stability at the glass transition of supercooled metallic liquids~\cite{Na2014} calls for deeper insights into the connection between the fragility index and the interatomic interaction.
As previous work suggested~\cite{Bennett1971,Hafner1987, Goetze2009}, 
mechanical stability in amorphous solids is crucially linked to the repulsive part of the interatomic interaction potential.
However, no consensus has been reached on whether interatomic repulsion \textit{softness} correlates with strong glasses~\cite{Mattsson2009} or with fragile glasses~\cite{Sengupta2012}. 
We derive an analytical closed-form relation between the fragility index of metallic glass formers and the short-ranged repulsive part of the interatomic interaction given by pseudopotential theory. 
This fundamental relation is obtained from a one-parameter theory fit to experimental rheological data of supercooled metallic melts. 
Resorting to this combination of theory and experiments, it is established that interatomic repulsion softness in metals goes along with strong glasses and low fragility. 
Surprisingly, given the difference in energy scale of many orders of magnitude and the nature of the microscopic interaction, this finding is in full agreement with the correlation observed experimentally for soft colloidal glasses by Mattsson, Weitz and co-workers~\cite{Mattsson2009}.
Finally, we establish a quantitative link between our analysis and the theory of shear transformation zones to estimate the size of the cooperatively rearranging regions in good agreement with 
the findings in Ref.~\cite{Liu2012}.
\section{Shear modulus of glasses}
The starting point for linking the shear modulus and the atomic connectivity analytically is the theoretical framework of nonaffine elastic response~\cite{Lemaitre2006,Zaccone2013,Dasgupta2013}.
The standard affine approximation of the classical Born-Huang theory is not applicable to amorphous as well as other non-centrosymmetric lattices~\cite{Alexander1998}.
This problem arises due to the lack of local inversion symmetry in amorphous solids. As a consequence, the deformation forces which are transmitted to an atom by its bonded neighbours do not balance each other by mirror symmetry.
The resulting forces, which act on every atom, are released through additional {\it nonaffine} motions on top of the standard {\it affine} displacements dictated by the macroscopic strain.
In other words, the continuum assumption that the macroscopic deformation scales down to the microscopic lattice does not generally hold for amorphous systems.

Structural disorder and nonaffine motions can be taken into account using the theory of nonaffine elastic response. 
For an amorphous solid under a shear strain $\gamma$, we can express the free energy of deformation as $F(\gamma)=F_{\text{A}}(\gamma)-F_{\text{NA}}(\gamma)$~\cite{Zaccone2014}.
The two terms represent the standard affine contribution to the free energy, provided by the framework of Born-Huang lattice dynamics~\cite{Born1954,Zaccone2011a}, and the nonaffine contribution, respectively.
Resorting to an eigenfunction decomposition of the nonaffine contribution, it is possible to derive an analytic expression for the shear modulus of an amorphous lattice.
This has been done for example by Lema\^itre and Maloney~\cite{Lemaitre2006} and the result for the shear modulus is given by
\begin{equation}\label{shearmod}
	G=G_{\text{A}}-G_{\text{NA}} 
	=G_{\text{A}}
	-
	\sum_{i,j} \underline{f}\vphantom{f}^{T}_{\, i} \; \underline{\underline{H}}\vphantom{H}^{-1}_{ij} \, \underline{f}\vphantom{f}^{}_{\, j},
\end{equation}
where $\underline{\underline{H}}\vphantom{H}_{ij} = (\partial^2 U/ \partial \underline{r}_i \partial \underline{r}_j )_{\gamma \to 0}$ represents the standard  dynamical matrix of the solid~\cite{ashcroft2005}, $U$ the internal energy of the system and $\underline{f}\vphantom{f}^{}_{\, i}$ the force acting on the atoms due to the shear deformation~\cite{Lemaitre2006}.
The explicit expression for the affine contribution to the shear modulus  is given by $G_{\text{A}} =(N/30V)\kappa R_0^2 Z$, which is discussed in the {\it Supporting Information}.
As shown in Ref.~\cite{Zaccone2013}, assuming a central-force interaction and introducing the atomic packing fraction $\phi = vN/V$, with $v$ a characteristic rigid-core volume, Eq.~\eqref{shearmod} can be evaluated analytically as
\begin{align}\label{nonaffine}
	G=G_{\text{A}}-G_{\text{NA}}=\dfrac{1}{5\pi} \dfrac{\kappa}{R_0}\phi (Z-Z_c).
\end{align}
The non-affinity of the amorphous solid is encoded in the quantity $-Z_c$, which denotes the critical number of bonds at which the shear modulus vanishes by  virtue of the nonaffine softening mechanism.
This expression still does not include the direct contribution of thermal effects to the elastic response. Thermal vibrations in fact soften the shear modulus by an additional negative term $-3(N/V)kT \partial^2( \ln \hbar\omega/kT)/\partial\gamma^2$~\cite{Zaccone2013, frenkel1955}.
For many materials, including metallic and polymer materials, this contribution is very small compared to the other terms in Eq.~\eqref{nonaffine}. It determines a decreasing trend of $G$ with $T$ which is negligible compared to the combined effect of non-affinity and thermal expansion~\cite{Luthi2007}.
\section{Temperature dependence of the shear modulus}	
The crucial effect which controls the temperature dependence of the shear modulus is the change in atomic connectivity $Z$ due to Debye-Gr\"uneisen thermal expansion~\cite{Zaccone2013}.
Approaching the glass transition temperature $T_g$ from below, this effect is responsible for the loss of mechanical stability.
We will show that the same effect is responsible for the decrease of the high-frequency shear modulus with increasing $T$ in the supercooled liquid above $T_g$.

The atomic packing fraction $\phi$ is reduced upon increasing the temperature $T$, an effect mediated by the thermal expansion coefficient defined as $\alpha_T = \frac{1}{V}(\partial V/ \partial T) = -\frac{1}{\phi}( \partial \phi/ \partial T)$.
Integrating this, we see that the atomic packing fraction evolves with $T$ according to $\log(1/\phi) = \alpha_T \, T + c$.
For an amorphous metal, a decrease in $Z$ arises if the separation between two particles is larger than the typical length scale of attraction defined by the first minimum of the interatomic pseudopotential $r_{\text{min}}$. For example, if the separation of an atom from one of its caged nearest-neighbours exceeds $r>r_{\text{min}}$, the neighbour  effectively leaves the coordination shell or cage~\cite{Zaccone2013}, and no longer contributes to the cage elasticity. 

\begin{figure}[t]
	\centering
	\includegraphics[width=.7\linewidth]{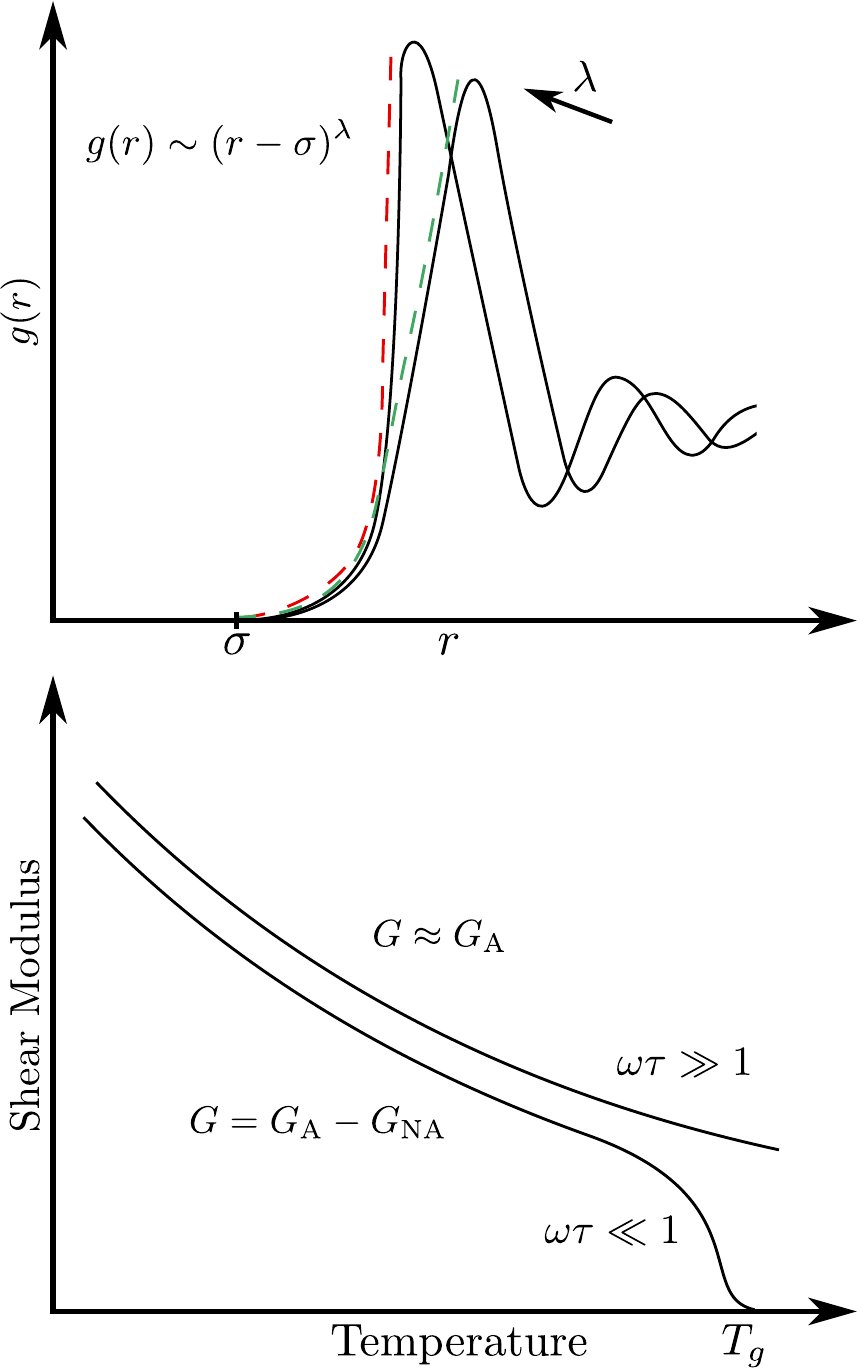}
	\put(-10,270){(a)}
	\put(-10,130){(b)}
	\caption{
		(a) Approximation of the repulsive part of the first peak of $g(r)$ using two different values for the steepness $\lambda$. An increase in $\lambda$ is linked to a steeper slope of $g(r)$.
		(b) In the high-frequency regime the affine shear modulus represents a good approximation to the actual behaviour of the shear modulus $G = G_{\text{A}} - G_{\text{NA}}$.				
	\label{fig1}}
\end{figure}

When increasing $T$, the average spacing between atoms in the coordination shell becomes larger, and the probability of nearest neighbours leaving the connectivity shell increases. 
It is then possible to use the radial distribution function $g(r)$ to relate the change in packing fraction $\delta \phi$, due to an externally imposed change in temperature $\delta T$, to the change in connectivity $\delta Z$.
Following along the lines of Ref.~ \cite{Zaccone2013}, the change of atomic connectivity $\delta Z = Z-Z_c$, relative to the critical stability (isostatic) point $Z_c$, can be calculated when the density of the system increases by an increment $\delta \phi = \phi - \phi_c$ according to
\begin{equation}\label{integral}
Z - Z_c \;  \sim \int_{1}^{1 +\delta \phi} r^2 g(r) \diff r,
\end{equation}	
where $r$ represents a dimensionless distance defined with respect to the rigid-core diameter $\sigma$. 
Since the radial distribution function $g(r)$ is not known in analytical form for real materials, we introduce an approximation scheme.
The basic idea is to represent the repulsive side of the first peak of $g(r)$ by means of the power-law approximation $g(r) \sim (r-\sigma)^{\lambda}$. 
In this way, the parameter $\lambda$ uniquely characterises the steepness of the left-hand side of the first-peak of the radial distribution function. The dashed lines shown in Fig.~\ref{fig1}~(a) represent the power-law approximation to the actual radial distribution function. 

We know that the potential of mean force~\cite{March2005} between two atoms is related to the radial distribution function by $V_{\text{m}}/kT = -\ln{ g(r)} \sim -\ln(r-\sigma)^{\lambda}$, where the ion-core diameter $\sigma$ indicates the mutual separation between two ions at which the interaction energy is practically infinite.
If the separation between two ions is small, $V_{\text{m}}$ reduces to the short-range part of the ion-ion repulsion.
Hence, $\lambda$ is proportional to the steepness of the short-range effective repulsion and inversely proportional to the softness of the pseudopotential, which scales as $1/\lambda$. 

Subsequently, with the power-law approximation for $g(r)$ in Eq.~\eqref{integral}, the change in connectivity becomes a function of the repulsion steepness $\lambda$:
$\delta Z \sim \delta \phi ^{1+\lambda}$.
It is assumed that the spherical integration is well approximated by Cartesian coordinates at short separations.
When decreasing the temperature by $\delta T<0$, the atomic packing fraction grows by $\delta \phi=-\phi\alpha_T \delta T>0$. Consequently, the connectivity $Z$ increases more strongly for steeper pseudopotentials than for the softer counterpart.
Analogously, an increase of temperature, $\delta T>0$, causes the atomic connectivity to decrease more abruptly with $T$ for a steep ion-ion repulsion, and more gradually for a softer interaction. 
\begin{table*}[t]
	\caption{Summary of the experimental data \label{tab1}}
	\begin{tabular*}{\hsize}
		{@{\extracolsep{\fill}}lrrrrrrc}
		\hline
		\multicolumn1l{\text{Alloy}}	&	\multicolumn1c{$T_g (K)$}		&		\multicolumn1c{$10^2 \alpha_T  T_g$}		&	\multicolumn1c{$\lambda$}		&	\multicolumn1c{$C_G \,(\text{GPa})$}		&		\multicolumn1c{$V_c (10^{-27}m^3)$}	&	\multicolumn1c{$m(\lambda)$}	 	&	\multicolumn1c{$m$}	 	\cr
		\hline
		Zr$_{46.75}$ Ti$_{8.25}$ Ni$_{10}$Cu$_{7.5}$  Be$_{27.5}$ & 597$^{\text{\cite{Johnson2007}}}$ & 0.591$^{\text{\phantom{\cite{Johnson2007}}}}$& 99.7& 34.52$^{\text{\phantom{\cite{Johnson2007}}}}$ &---&---&---\cr
		Pd$_{43}$ Cu$_{27}$ Ni$_{10}$ P$_{20}$ & 567$^{\text{\cite{Johnson2007}}}$ & 0.935$^{\text{\cite{Kato2008}}}$&115.9
		&30.44$^{\text{\phantom{\cite{Johnson2007}}}}$ &---&---&---\cr
		Pt$_{57.5}$ Ni$_{5.3}$ Cu$_{14.7}$ P$_{22.5}$ & 489$^{\text{\cite{Johnson2007}}}$ &0.776$^{\text{\phantom{\cite{Johnson2007}}}}$   & 164.2& 30.56$^{\text{\phantom{\cite{Johnson2007}}}}$&---&---&---	\cr		
		La$_{55}$ Al$_{25}$ Ni$_{20}$ & 465$^{\text{\cite{Kato2008}}}$ & 0.711$^{\text{\cite{Kato2008}}}$ &196.2		&15.4 $^{\text{\cite{Wang2011}}}$&0.0148&37.16	&37$_{450K}^{\text{\cite{Johnson2007}}}$$\quad$ 33$_{462K}^{\text{\cite{Okumura1993}}}$\cr
		Zr$_{41.2}$ Ti$_{13.8}$ Ni$_{10}$ Cu$_{12.5}$ Be$_{22.5}$ & 623$^{\text{\cite{Kato2008}}}$ & 0.617$^{\text{\cite{Kato2008}}}$&276.4&33.2$^{\text{\cite{Johnson2007}}}$&0.0085 &38.74&40$_{613K}^{\text{\cite{Johnson2007}}}$$\quad$ 39$_{648K}^{\text{\cite{Waniuk1998a}}}$	\cr		
		Pd$_{40}$ Ni$_{40}$ P$_{20}$ & 551$^{\text{\cite{Kato2008}}}$ & 0.856$^{\text{\cite{Kato2008}}}$&286.5&	36.5 $^{\text{\cite{Wang2011}}}$&0.0069&49.91&50$_{560K}^{\text{\cite{Johnson2007}}}$$\quad$ 41$_{580K}^{\text{\cite{Chen1978}}}$	\cr
		Pd$_{77.5}$ Cu$_{6}$ Si$_{16.5}$  & 625$^{\text{\cite{Kato2008}}}$ & 0.865$^{\text{\cite{Kato2008}}}$&381.2&32.9 $^{\text{\cite{Wang2011}}}$&0.0084 &60.04&61$_{634K}^{\text{\cite{Johnson2007}}}$$\quad$ 52$_{635K}^ {\text{\cite{Chen1978}}}$	\cr	
		\hline
	\end{tabular*}
\end{table*}

\section{The high-frequency shear modulus}
Experimental measurements of the viscosity and shear modulus of supercooled liquid metals at the glass transition can be obtained using ultrasonic techniques, which probe the material response at frequencies in the GHz range~\cite{Johnson2007}. 
These high frequencies exceed the typical relaxation frequency of a metallic glass by several orders of magnitude~\cite{Wang2011}.
Under such conditions, the response to an applied oscillatory shear strain generally is dominated by the instantaneous (affine or quasi-affine) limit of the shear modulus. 
For frequencies $\omega$ much larger than the inverse of the Maxwell relaxation time $\tau$, that is for $\omega \tau \gg 1$, the shear modulus cannot decay through a nonaffine relaxation process.
This situation is sketched in Fig.~\ref{fig1}~(b) together with the low-frequency case, where the nonaffine decay is possible.

At high frequency, the atoms cannot leave their affine positions to reach the nonaffine positions, as the deformation is too quickly reverted.
Hence, the elastic response at GHz frequencies is predominantly affine, as shown in the {\it Supporting Information} using the full nonaffine response theory. 
Considering Eq.~\eqref{nonaffine}, this means that the shear modulus is reduced to its affine contribution in the sense that $G \xrightarrow{\omega \to \infty}  G_{\text{A}}$.
Consequently, in this regime the expression for $ G$ is proportional to $Z$~\cite{Born1954}, but no longer depends on the critical connectivity $Z_c$. Therefore, it holds true that $\delta Z \to Z$ and $\delta \phi \to \phi$. Setting $Z_c$ and $\phi_c$ to zero is the defining feature of the high-frequency quasi-affine limit \cite{Zaccone2011a}.

Going back to Eq.~\eqref{shearmod}, we recognise that in the regime $\omega \tau \gg 1$ this leaves us with $G = \frac{1}{5 \pi} \frac{\kappa}{R_0} \phi Z $.
We recall that the packing fraction depends on $T$, $\phi(T) \sim \mathrm{e}^{-\alpha_T T}$ and, thus, we obtain  $Z(T) \sim\mathrm{e}^{-(1+\lambda)\alpha_T T} $. Upon replacing this result in the above equation for $G$, we find that the $T$-dependence of the shear modulus is dictated by
\begin{align}\label{shearpotential}
G(T) \sim \dfrac{1}{5 \pi} \dfrac{\kappa}{R_0}
\exp\Big[ -(2+\lambda) \alpha_T T \Big].
\end{align}
The high-frequency shear modulus now explicitly depends on the softness of the interaction potential, and on the thermal expansion coefficient $\alpha_T$. Both these crucial effects are reflections of the \textit{anharmonicity} of the elastic response.

As already hinted above, we shall remark that in general there is also a phonon contribution to the shear modulus proportional to $kT \mathrm{e}^{-\alpha_T T}$. However, this contribution is typically negligible with respect to the one in Eq.~\eqref{shearpotential}~\cite{Luthi2007}, even more so, if one considers, as it will be shown below, that typical values of $\lambda$ are in the range $100-400$. 
\section*{Comparison with experimental data}
\begin{figure}[b]
	\centering
	\includegraphics[width=.9\linewidth]{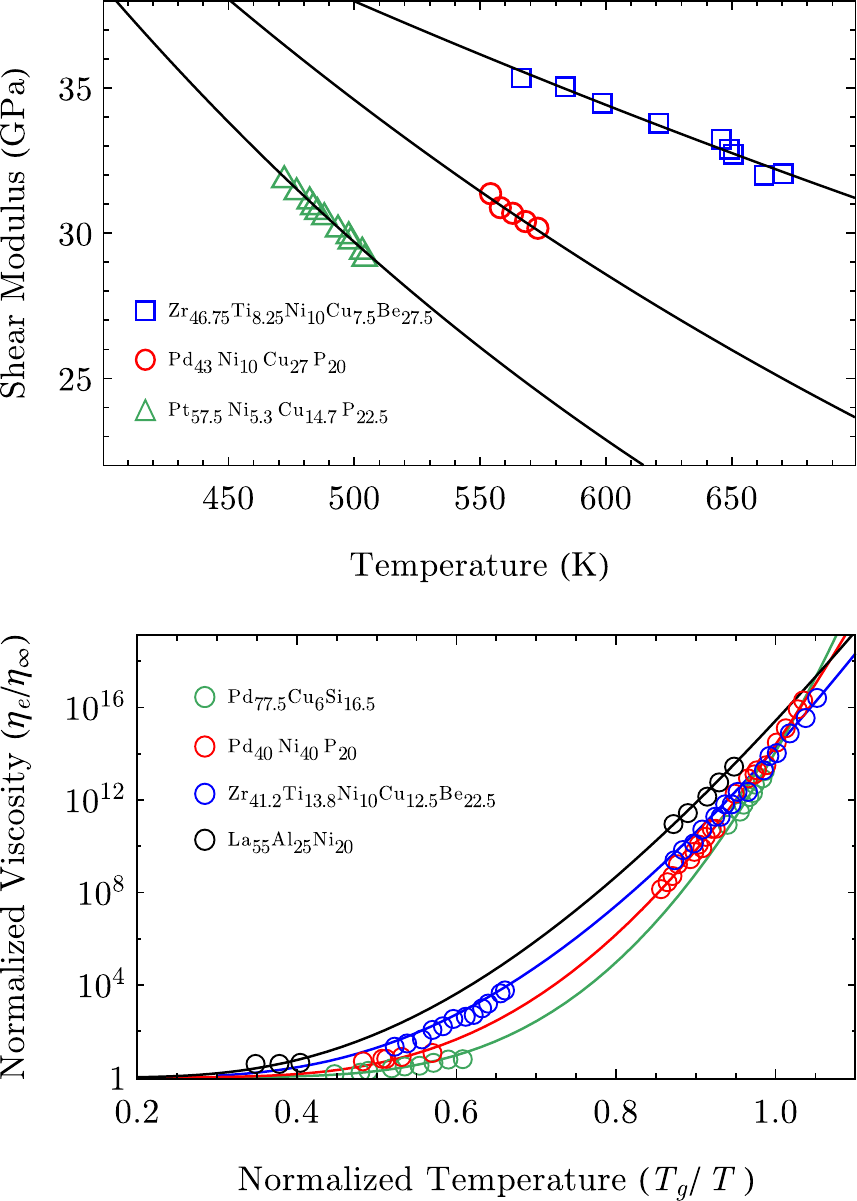}
	\put(-18,220){(a)}
	\put(-18,43){(b)}
	\caption{The experimental data points for various glass-forming alloys from \cite{Johnson2007} and the respective fitting curves for the shear modulus in panel (a), and the viscosity in panel (b). The solid lines are the one-parameter fitting curves obtained using the expressions Eqs.~\eqref{eq_th_shear} and~\eqref{thvisc}, for the shear modulus and viscosity, respectively.  The values used for the fittings are reported in Tab.~\ref{tab1}.
		\label{Viscfit}
	}
\end{figure}

The above expression for the high-frequency affine shear modulus can be rewritten as
\begin{align}\label{eq_th_shear}
G(T)
=
C_G
\,
\exp{\left[		\alpha_T	 T_g (2+\lambda)	\left(1-\frac{T}{T_g}\right)	\right]},
\end{align}
where $C_G=\frac{\varepsilon}{5\pi}	\frac{\kappa}{R_0}\mathrm{e}^{-\alpha_T T_g (2+\lambda)} $ is a prefactor independent of $T$. The constant $\varepsilon$ stems from the integration of $\alpha_T$ and from the dimensional prefactor in the power-law ansatz for $g(r)$. All the parameters in this expression, which are given in Tab.~\ref{tab1}, are fixed by the experimental protocol, apart from the fitting parameter $\lambda$ related to the ion-ion repulsion steepness.
With Eq.~\eqref{eq_th_shear} at hand, we can generate a one-parameter fit to the experimental data provided from Ref.~\cite{Johnson2007}, which accurately captures the data sets for the three metallic glass alloys, as can be seen from Fig.~\ref{Viscfit}~(a). The different slope of the three depicted curves reflects the fact that the repulsion steepness $\lambda$ in Eq.~\eqref{eq_th_shear} controls the behaviour of $G(T)$. A decreasing $\lambda$, among the different alloys, correlates with a slower decrease of the shear modulus upon increasing the temperature.

Furthermore, we can use our model for the high-frequency shear modulus to evaluate the activation energy $E(T)$ involved in restructuring the glassy cage and, hence, the viscosity $\eta$ of the melts. Within the framework of the shoving or elastic model of the glass transition~\cite{Eyring1943, Eyring1936, Dyre1998,Dyre2006}, the activation energy for local cooperative rearrangements is $E(T)=G_{\text{A}} V_{\text{c}}$. The characteristic atomic volume $V_{\text{c}}$ appearing here is accessible through the theoretical fitting to the viscosity data, although its value is approximately specified by the atomic composition of the alloy.

Replacing the expression for the activation energy in the Arrhenius relation given by the shoving model of the glass transition, and using Eq.~\eqref{eq_th_shear} for the high-frequency shear modulus $G_{\text{A}}$ inside $E(T)$, we obtain the following analytical expression for the viscosity, 
\begin{align}\label{thvisc}
\dfrac{\eta(T)}{\eta_{0}}
=
\exp{\left\{ \dfrac{V_c C_G}{k \, T}		\exp{\left[		(2+\lambda)	\alpha_T T_g \left(	1-\frac{T}{T_g}\right)	\right]}		\right\}},
\end{align}
where $\eta_0$ is a normalisation constant.

It is important to note how the \textit{double-exponential} of the viscosity versus $T$ arises. The first exponential stems from the elastic activation described in the framework of the shoving model, whereas the second exponential is due to the Debye-Gr\"uneisen thermal expansion rooted in lattice-dynamical
considerations, and ultimately related to anharmonicity.

We compare the theoretical predictions to the experimental data of Ref.~\cite{Johnson2007} in Fig.~\ref{Viscfit} (b). In this case there is also an excellent agreement between theory and experiment with the adjustable parameters being $\lambda$, the steepness of the short-ranged ion-ion repulsion, and $V_{\text{c}}$, the characteristic atomic volume.
\section{Interatomic repulsion and fragility}
With the analytical theory developed above, we are in the position to relate the atomic-scale properties of the interaction between ions to the experimentally observable macroscopic response of the material.
We now consider the behaviour of the viscosity in Fig.~\ref{Viscfit} (b) together with the corresponding behaviour of the interaction parameter $\lambda$ for various alloys in Tab.~\ref{tab1}.
Evidently, upon approaching the glass transition, the slope of the viscosity $\eta(T)$ is controlled by the interatomic repulsion steepness $\lambda$, which depends on the atomic composition of the alloy.
A steeper pseudopotential repulsion between two nearest-neighbour ions goes hand in hand with a steeper rise of viscosity, when $T$ is increased. 

This observation leads us straight to connecting the softness of the potential to the fragility of metallic glasses. The fragility is given as the slope of the viscosity evaluated at the glass transition temperature $T_g$, i.e.
$m= \big(\frac{\partial \log_{10}(\eta / \eta_0)}{\partial (T_g/T)}\big)\big \vert_{T=T_g}$~\cite{Angell2000}.
Using the analytical expression for $\eta$, Eq.~\eqref{thvisc}, we obtain a simple relation between the fragility $m$ and the steepness of the interatomic repulsion $\lambda$ given by
\begin{align}
m(\lambda)
=
\dfrac{1}{\ln10}
\dfrac{V_c C_G}{k T_g}
\Big[
1+ (2+\lambda) \alpha_T T_g
\Big].
\end{align}
Metallic glasses with a steeper repulsive part of the interatomic interaction are thus more fragile.
The values of the fragility obtained for the various alloys of Ref.~\cite{Johnson2007} are listed in Tab.~\ref{tab1}, together with the fitted values of the interatomic repulsion steepness $\lambda$. Good agreement is also found with independent experimental measurements of $m$ from the literature.

Intriguingly, this prediction is in full agreement with the experimental findings of Ref.~\cite{Mattsson2009}. In that work  the softness of the interparticle potential was varied in a model colloidal glass, where the energy scale is orders of magnitude smaller than in metals.  

The model also can capture the behavior of $m$ observed in simulation studies of Lennard-Jones glasses, where the attractive anharmonicity controls the fragility via the thermal expansion coefficient and, by construction, a high anharmonicity is accompanied by a low repulsion steepness ~\cite{Bordat2004}.

\section{Extracting pseudopotentials from experimental data}
Given the schematic form of the repulsive short-range part of the interaction, $-\ln(r-\sigma)^{\lambda}$, used in the fitting, it is desirable to map this semi-empirical repulsion onto a physically realistic interatomic pseudopotential. 
This can be achieved by using an Ashcroft-type pseudopotential for modelling the Thomas-Fermi screened interionic Coulomb repulsion~\cite{Finnis2003} and, in addition, a Born-Mayer interaction term which accounts for the effect of electron-overlap and Pauli exclusion repulsion between valence-electron shells of two interacting ions~\cite{Mott1958}. 
A more detailed discussion of this matter can be found in the {\it Supporting Information}.
This combination of the two contributions to the interaction is the most meaningful choice for the present situation, as discussed in~\cite{Torrens1972}.

The softness of the pseudopotential is predominantly controlled by the Born-Mayer parameters, because electron-overlap repulsion between valence electrons is more energetic over a broader length scale compared to the Ashcroft contribution, as illustrated in the {\it Supporting Information}. Physically, a slower decay of the electron-overlap repulsion with distance reflects the softness of the effective interaction. The glass stability, however, is optimised by the coexistence of both softness and substantial repulsion, as is the case for technologically important alloys, like binary Zr-Cu alloys~\cite{Samwer1982}.

\begin{figure}[t]
	\centering
	\includegraphics[width=.8\linewidth]{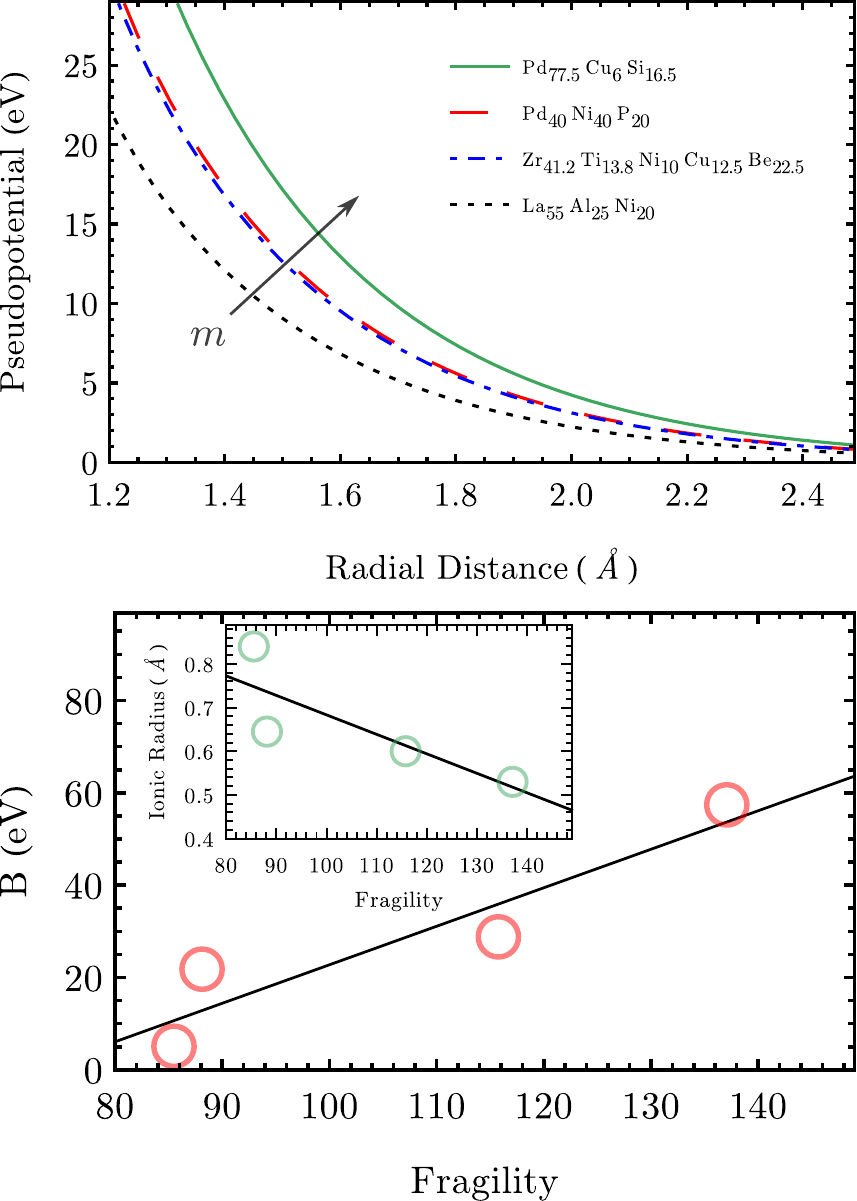}
	\put(-18,267){(a)}
	\put(-18,125){(b)}
	\caption{
		(a) The Ashcroft-Born-Mayer pseudopotential is depicted for four different glass-forming alloys. The fragility $m$ increases with the pseudopotential steepness.
		(b) The value of the Born-Mayer energy scale increases linearly with the fragility. Also, it is observed that the average ionic diameter decreases linearly with the fragility.			\label{pseudoplot}}
\end{figure}

Microscopically, it is the strongly anisotropic density distribution of d-shell electrons, due to the quadrupolar d-wave symmetry, which provides significant softness (upon taking a spherical average), compared to the more isotropic electron density distribution of elements whose outer shells are dominated by s-electrons.
Hence, the form of the pseudopotentials may explain the difference in stability and fragility based on the composition of the alloy.
In our model, this effect is expressed by the energy-scale of the Born-Mayer repulsion $B$. In particular, we find that $B$ correlates linearly with the fragility index $m$, as shown in Fig.~\ref{pseudoplot}~(b). This correlation reflects the fact that d-shell orbitals effectively soften the interatomic repulsion, whereas s-shell electrons are associated with steeper repulsion and higher fragility. 

The second effect which is captured by this approach is the ion-size mismatch. If smaller metal atoms are added to larger atoms, fragility decreases and strong glasses can be formed. This mechanism which affects multi-component alloys is analysed and discussed in~\cite{Egami2002}.  Again, this is the consequence of an effectively softer interatomic repulsion.
Smaller atoms of metalloids like P, B or Si can easily come closer to larger ions like Pd, La, Zr or Cu by fitting into the``voids" of the quadrupolar d-shell structure. In general, this topological effect also leads to a softer average pseudopotential.

This connection between macroscopic flow behaviour, encoded in $m$, and electronic structure is an important step towards a unifying framework for understanding and controlling mechanical properties of metallic glasses on the atomic scale. 
\section{Connection with cooperative shear events}
As already pointed out, the energy necessary to trigger a shoving event is
$
E(T) = G_{\text{A}}(T) V_c.
$
Importantly, the characteristic atomic volume $V_c$ is not the volume change which is connected to a shoving event. This quantity, also called the activation volume $\Delta V$, is connected to $V_c$ and the initial shoving volume $V$ via the relation
\begin{equation}\label{shovingmodel}
V_c = \dfrac{2}{3} \dfrac{(\Delta V)^2}{V},
\end{equation}
which can be derived in the framework of the elasticity theory of an isotropic expanding sphere~\cite{Dyre1998}.

It is widely believed that shear transformation zones (STZs) are the fundamental plastic entities responsible for the yielding mechanism in metallic glasses. STZs are clusters of atoms which can cooperatively rearrange under shear stress, and are directly connected to the local accumulation of free volume~\cite{Li2013a}. It is in this sense that the activation of STZs allows the involved atoms to rearrange more easily under shear stress. Assuming that the initial shoving volume $V$ corresponds to the volume of a STZ, we find a direct relation between the characteristic volume $V_c$ and the activation volume $\Delta V$ in the following way.

According to Ref.~\cite{Johnson2005}, the total energy barrier $W$ between two basins in the potential energy landscape
can be evaluated to give $W \approx (1/320) \,G_{\text{A}} \Omega$, see {\it Supporting Information}.
We assume that $W$ in the cooperative shear model is approximately equal to the shoving energy, that is $W \approx E$. 
It directly follows that $G_{\text{A}} V_c \approx (1/320) G_{\text{A}} \Omega$, which leads us to conclude that the effective volume of a STZ is  $\Omega \approx 320 V_c$. 

It is physically meaningful that the effective STZ volume $\Omega$ is approximately equivalent to the initial shoving volume $V$, henceforth calling it $V_{\text{STZ}}$.
Using the identification $\Omega \approx V_{\text{STZ}}\approx320 V_c$, we can use the values for $V_c$ to extract values for STZ volumes from our theoretical analysis and compare them to experimental results for $V_{\text{STZ}}$ from Ref.~\cite{Liu2012}. We find that the calculated STZ volumes for the respective alloys are in very good agreement with the experimental results for similar alloys, which is displayed in Tab.2. 

Moreover, using $V_{\text{STZ}}\approx320 V_c$
together with Eq.~\eqref{shovingmodel}, we obtain a relation  between the activation volume and the characteristic volume $V_c$ given by $\Delta V \approx \sqrt{480} \,V_c$. 
With the values for $V_c$ from the viscosity fitting, we can calculate the activation volume for the corresponding alloys to be in the range $151 - 324 \, \mathring{A}^3$.
For a Pd-based metallic glass, an activation volume of $106 \mathring{A}^3$ was found experimentally~\cite{Heggen2005}, which is not too far from our estimate for the alloys discussed here. 

Elsewhere the activation volume for Zr$_{41.2}$ Ti$_{13.8}$ Cu$_{12.5}$ Ni$_{10}$ Be$_{22.5}$ is determined to be $75 \,\mathring{A}^3$~\cite{Schuh2007}. For the same alloy, we calculate the value for the activation volume from the corresponding $V_c$ with the result $\Delta V=186 \, \mathring{A}^3$, which is about $2.5$ times larger in comparison. This difference may be explained by the different deformation protocols (shear amplitude, applied stress rate, etc.), 
in the respective experiments. It is argued in Refs.~\cite{Fan2014,Zhang2015} that a higher degree of applied stress leads to an increase of the size of the individual flow units, which means that both the STZ volume and the activation volume tend to increase.
\begin{figure}[t]
	\centering
	\includegraphics[width=1\linewidth]{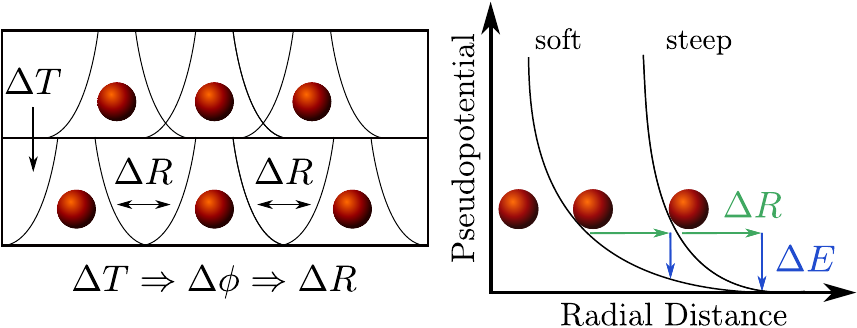}
\caption{
(Left) 
 The distance between the atoms decreases as the temperature is increased leading to a smaller overlap of the effective interaction potentials.
(Right)
The growth of the cage by $\Delta R$ when increasing the temperature by $\Delta T $ and the corresponding loss of stabilizing energy $\Delta E$. The potentials are shifted for the sake of clarity.	
		 \label{deltae}}
\end{figure}%
\begin{table}[h]\label{tab2}
	\caption{Experimental and theoretical STZ volumes}
	\begin{tabular}{@{ \vrule height 10.5pt depth4pt  width0pt}lll}
		\hline
		Alloy &$V_c \, (\text{nm}^3)$ &$\Omega \, (\text{nm}^3)$
		\\
		\hline
		Pd$_{40}$ Ni$_{40}$ P$_{20}$
		&0.0069&	2.21 \\

		Pd$_{48}$ Ni$_{32}$ P$_{20}$
		&-&	2.36$^{\text{\cite{Liu2012}}}$ \\

		Zr$_{41.2}$ Ti$_{13.8}$ Ni$_{10}$ Cu$_{12.5}$ Be$_{22.5}$
		&0.0085	&2.72		\\

		Zr$_{46.75}$ Ti$_{8.25}$ Ni$_{10}$ Cu$_{7.5}$ Be$_{27.5}$
		&-	&3.13$^{\text{\cite{Liu2012}}}$		\\
		
		La$_{55}$ Al$_{25}$ Ni$_{20}$
		&0.0148	&	4.74	\\

		La$_{55}$ Al$_{25}$ Ni$_{20}$
		&-	&	5.31$^{\text{\cite{Liu2012}}}$	\\
		\hline
	\end{tabular}
\end{table}

\section{Conclusions}
The basic mechanism controlling the mechanical response and the fragility of liquid metals close to vitrification can be summarised in the following way.
Due to thermal expansion, an increase in the temperature leads to a decreasing atomic packing fraction and, thus, to a decrease of atomic connectivity.

The latter effect softens the material, causing the shear modulus to decrease with $T$. 
The rate of this process is controlled by the steepness of the repulsive short-range interatomic interaction. 
This mechanism propagates to the viscosity, and it controls its temperature dependence
and leads to fragile behaviour with steep interatomic repulsion, and to strong glasses when the repulsion is softer. 
In an amorphous solid we can picture this situation by considering a reference atom which is surrounded by a number of neighbouring atoms, forming a disordered cage. 
The repulsive interaction between these particles provides stability to the cage.
When the temperature is increased, a corresponding change of packing fraction takes place, implying that the disordered cage around the reference atom becomes larger and less stable, see Fig.~\ref{deltae}.
 
With this moving farther apart of the nearest neighbours from the reference atom, the local stabilising energy felt by the atoms decreases due to a smaller overlap of the repulsive interatomic interactions by $\Delta E$.
At the onset of the glass transition, the stabilising effect of the atomic cage breaks down, which ultimately leads to the vanishing of the zero-frequency shear modulus at $T_g$.
It is the steepness of the repulsive pseudopotential which controls how rapidly or abrupt the the stabilizing energy decreases as the temperature is increased. For an alloy whose constituents exhibit a steeper interatomic repulsion, this process of destabilisation will be more abrupt resulting in a faster variation of the shear modulus and viscosity with $T$ and, correspondingly, to a more fragile glass.

We also show that the steepness of the interatomic repulsion for various metallic alloys can be mapped one-to-one onto a pseudopotential with two contributions. The overall softness of the pseudopotential is mainly controlled by Born-Mayer repulsion stemming from the overlap of valence-shell electrons.
A direct relation of linear proportionality between the fragility index $m$ and the Born-Mayer energy $B$ is obtained from the fitting to experimental data. Lower values of $B$ may correlate with mixtures of elements having outer electrons in d-shells, as is the case of Cu in Zr-Cu alloys, or with the concentration of metalloid in metal-metalloid mixtures. Systematic studies in the future using ab-initio simulations may shed light on the link with the detailed electronic structure.

Furthermore, we connect the characteristic atomic volume $V_c$ with the size of STZs. In this regard, STZs appear to be regions in the amorphous solid with a relatively low average atomic connectivity $Z$. 
These regions are prone to elastic stress accumulation, leading to  an increase in individual shoving events, which eventually results in macroscopic plasticity.

While there exists a clear linear relation between the fragility and the repulsive steepness $\lambda$, the correlation between the size of a STZ and the fragility exhibits no simple form and remains to be understood in future investigations. 
We believe that the present framework may open up the possibility, in future work, of a priori designing metallic glasses with tailored rheological and mechanical properties (e.g. plasticity and ductility~\cite{Kumar2013a} based on the alloy elemental composition.
\begin{acknowledgments}
The support of the 
Technische Universit\"at 
M\"unchen - Institute for Advanced Study, funded by the German Excellence Initiative and the EU 7$^{\text{th}}$ Framework Programme under Grant Agreement No. 291763 is acknowledged.

\end{acknowledgments}

\newpage

\section{{Supporting Information}}
\vspace{.7cm}
\subsection{The affine shear modulus}
The expression for the affine shear modulus $G_{\text{A}}$, Eq. \textbf{[1]} in the main text, can be obtained in the following way~\cite{Born1954, Zaccone2011b}. 
An expansion of the free energy, where the macroscopic strain is scaled down to the atomic displacements, yields the famous Born-Huang formula for the shear modulus in the harmonic approximation as $G_{\text{A}} = (1/V) \kappa R_0^2 \sum_{\langle i,j \rangle  }n_{ij}^x n_{ij}^yn_{ij}^xn_{ij}^y$. Here, $\kappa$ denotes the spring constant of an harmonic bond, and $R_0$ the equilibrium lattice constant. The bond-orientation unit vector $\underline{n}\vphantom{n}_{\, ij}= 	(	\cos \varphi \sin \theta, \sin \varphi \sin \theta , \cos \theta	)$ connects the nearest-neighbour atoms $i$ and $j$. In a lattice of $N$ atoms every bond contributes a term proportional to the geometric factor $\langle  n_{ij}^x n_{ij}^yn_{ij}^xn_{ij}^y\rangle$ to the above lattice sum. This average is to be evaluated for a prescribed angular distribution of the bond orientations \cite{Zaccone2011b}.  Introducing the average number of mechanical bonds per atom $Z$ (atomic connectivity), the mean-field estimate gives $G_A = (N/2V) \kappa R_0^2 Z \langle  n_{ij}^x n_{ij}^yn_{ij}^xn_{ij}^y\rangle$.
Assuming an isotropically random distribution of bond orientations, the average over the components of the bond orientation vectors yields the numerical factor $1/15$. Thus, we can establish an explicit expression for the affine shear modulus  as $G_{\text{A}} =(N/30V)\kappa R_0^2 Z$.

\subsection{The High-Frequency Shear Modulus in the Continuum Limit}

Following the exposition in \cite{Lemaitre2006}, the evaluation of the complex version of the nonaffine contribution to the shear modulus in  Eq. \textbf{[1]}  yields
\begin{align}
	G ^{\ast}(\omega)=
	G_A
	+
	\dfrac{1}{V}
	\sum_p
	\dfrac{\widehat{\Xi}_p \;\widehat{\Xi}_p}{m \omega ^2 - m \omega^2_p - i \omega \nu},
\end{align}
where $\widehat{\Xi}_p$ denotes the affine force field tensor and $\omega_p$ the eigenfrequencies of the system.
In the continuum limit we can replace the sum over the $p$-states by an integration over the eigenfrequencies $\omega_p$, that is
\begin{align}
	G ^{\ast}
	(\omega)
	=
	G_{\text{A}}
	+
	\int^{\omega_{\text{D}}}_0
	\diff \omega_p
	\dfrac{\rho(\omega_p) \Gamma(\omega_p)}{m \omega^2 - m \omega_p^2 - i \omega \nu},
\end{align}
where $\rho(\omega_p)$ denotes the density of states and $\omega_D$ the Debye frequency. The function $\Gamma$ represents the correlator on frequency shells given by $\Gamma (\omega) = \langle\widehat{\Xi}_p \,\widehat{\Xi}_p \rangle$, where the the average is evaluated for all projections of $\Xi$ on eigenvectors with eigenfrequency $\omega_p \in [\omega, \omega + \diff \omega]$.

We can express the shell correlator in a simple fashion due to the result in \cite{Zaccone2011a} as $\Gamma \sim \omega_p^2$. Also we will assume that the density of states is quadratic as in the Debye model given as $\rho(\omega_p) \sim \omega_p^2$. 

Therefore the expression for the complex shear modulus reduces to
\begin{align}
	G ^{\ast}
	(\omega)
	=
	G_{\text{A}}
	+
	\int^{\omega_{\text{D}}}_0
	\diff \omega_p
	\dfrac{\omega_p^4}{m \omega^2 - m \omega_p^2 - i \omega \nu}.
\end{align}
Since we are interested in the shear modulus it suffices to consider the real part of the complex quantity $G^{\ast}$. The real part of the integrand above is easily seen to be
\begin{align}
	\dfrac{2 m \omega_p^4\left(\omega^2 - \omega_p^2\right)}
	{m^2 \omega^4\left(1+ \dfrac{\omega_p^4}{\omega^4}-2\dfrac{\omega_p^2}{\omega^2}\right)}.
\end{align}
Thus, we observe, that in the high-frequency limit, where $\omega \to \infty$, the above expression will converge to zero, leading to $\lim_{\omega \to \infty} \text{Re} \; G^{\ast}(\omega)=G_A$. By virtue of this argument we can substitute the high-frequency shear modulus with the Born-Huang affine shear modulus $G_A$.

\subsection{The Ashcroft-Born-Mayer potential}

With the two contributions introduced in the main text, the Ashcroft-Born-Mayer pseudopotential is given by
\begin{align}
V_{\text{ii}}(r)
=
\dfrac{A \, \mathrm{e}^{-q_{\text{TF}}(r-2a_0)}}{r-2a_0
	}
	+
	B\,
	\mathrm{e}^{ -C(r-\bar{\sigma})},
	\end{align}
where $a_0$ is the Bohr radius and $\bar{\sigma}$ the average ionic core diameter of the alloy, which corresponds to the average size of the ionized atoms constituting the alloy. The average ionic core diameter is obtained by a averaging the respective ionic core diameter of the constituents with their contributing weights given by their volume ratios in the alloy. The values for the ionic core diameters of the atoms constituting the alloys are taken from Ref.~\cite{Shannon1976}. The quantities $A$ and $B$ set the energy scales for the repulsive interaction from the Ashcroft and Born-Mayer term, respectively.
The parameter $q_{\text{TF}}$ is the inverse of the Thomas-Fermi screening length given by Thomas-Fermi theory, and its value is known for different types of alloys~\cite{Wang2004}. We choose a representative value for $q_{\text{TF}}$ as $1.25 \, \mathring{A}^{-1}$  according to the values reported in Ref.~\cite{Wang2004}. The ionic core diameter $\bar{\sigma}$ is obtained by a weighted average of the core diameters of the atoms constituting the alloys taken from~\cite{Shannon1976}, where the weights correspond to the ratios of the respective atoms.

The characteristic range $1/C$ of the valence-shell overlap repulsion is not known a priori. However, its typical values are less sensitive to the atomic composition than the parameters $\bar{\sigma}$, $A$ and $B$. Different atoms have very similar values typically in the range $1-2.5 \, \mathring{A}^{-1}$~\cite{Nikulin1970}. 

Finally, $A$ is the prefactor to the Ashcroft pseudopotential, which is the product of the electrostatic nuclear repulsion $Z_{\text{ion}}^2 e^2$ and the Ashcroft factor $\cosh^{2}(q_{\text{TF}}  R_{\text{core}})$~\cite{Faber1972}, where $ R_{\text{core}}$ is a typical value for the atom-specific core radius and  $Z_{\text{ion}}$ the effective ionic charge number. The latter cannot be easily estimated from first-principles or from literature. Similarly, the prefactor $B$ of the Born-Mayer term, can be rigorously evaluated only from the exchange integrals of the various overlapping electrons which belong to the valence shells of the two interacting ions. This calculation, even in approximate form, is not feasible except for simple crystals. Hence, we take both $A$ and $B$ as adjustable parameters in the mapping between our schematic logarithmic potential and the Ashcroft-Born-Mayer pseudopotential. We shall remark that the Born-Mayer prefactor $B$ typically has non-trivial large variations from element to element across the periodic table, as shown in many ab initio simulations studies~\cite{Nikulin1970, Hafner1987}. Consistent with this known fact, it turns out that $B$ is the most sensitive parameter in our analysis, in the sense that small variations in $B$ can lead to large deviations in the fitting of the experimental data. Conversely, the Ashcroft prefactor $A$ is not a sensitive parameter, and its values are not crucial for the match with experiments. 

\begin{figure}[b]
	\centering
	\includegraphics{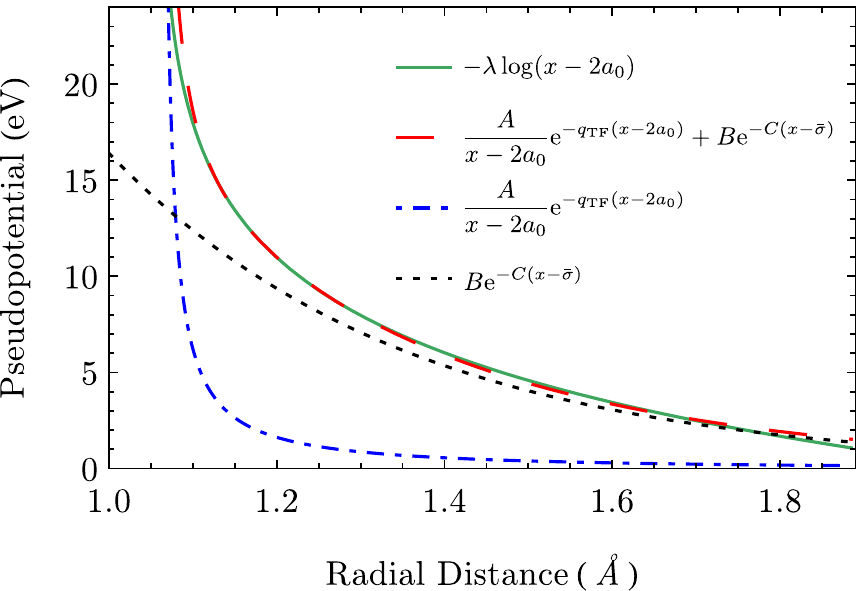}
		\caption{Comparison of the Ashcroft-Born-Mayer pseudopotential with the logarithmic potential of mean force (including the two separate contributions to the pseudopotential). The plot was generated for a repulsive steepness $\lambda = 99.7$.\label{potentialcomparison}}
\end{figure}

\subsection{Pseudopotential Fit}
The values of the parameters which have been used to obtain fits to the logarithmic pseudopotential are documented in Table 1 of this {\it Supporting Information}. 

In Fig.~\ref{potentialcomparison} we plotted a comparison between the Ashcroft-Born-Mayer pseudoptential and the potential of mean force. We see that the repulsive form of the potential of mean force is well approximated by the exponential functional form of the pseudopotential.

The values for both the Thomas-Fermi wave vector $q_{\text{TF}}$ and the Pauli exclusion parameter are seen to vary very little for the different alloys. In contrast, the parameters which define the energy scale of the two contributions to the Ashcroft-Born-Mayer potential show a significant variation. Interestingly, this variation correlates with the steepness parameter $\lambda$ defined earlier. In particular, a increasing values for the steepness $\lambda$ coincide with an increase in the energy scale, which is defined by both the parameters $A$ and $B$.

Also, we can deduce from Fig.~\ref{potentialcomparison}  and Tab.~\ref{potentialfitting} that the Born-Mayer term in the pseudopotential provides the dominant contribution to the repulsive effective potential at longer length scales. The screened Coulomb interaction represented by the Ashcroft term only comes into action at distances shorter than the screening length, where the electrostatic repulsion of the ionic core provides a harsh repulsive barrier.

\subsection{Comparison of the Born-Mayer parameter with typical values}
Our modelling of $\lambda$ indicates clearly that, of the two contributions, it is the Born-Mayer electron-overlap repulsion the one which controls the qualitative behaviour, while the A parameter does not have much effect on the slope. This implies that the parameters of the Born-Mayer repulsion are those which control the interaction softness. Furthermore, if we recast the Born-Mayer interaction in the canonical form which is found in the molecular physics literature, $Be^{-Cr}$, for our Zr-based alloy we obtain  $B\approx 15000~eV$ which is not far from the value $B=22000~eV$ obtained in simulations of the interaction between two Zr atoms in the literature, e.g. Ref.~\cite{Abrahamson1969}. This comparison confirms the robustness of our model, not only at the qualitative level, but also in capturing the right orders of magnitude of the microscopic forces. We believe that more detailed calculations using ab-initio methods can be done along this line in future work by ab-initio molecular dynamics specialists.
\begin{figure}[t]
	\centering
	\includegraphics[width=0.7\linewidth]{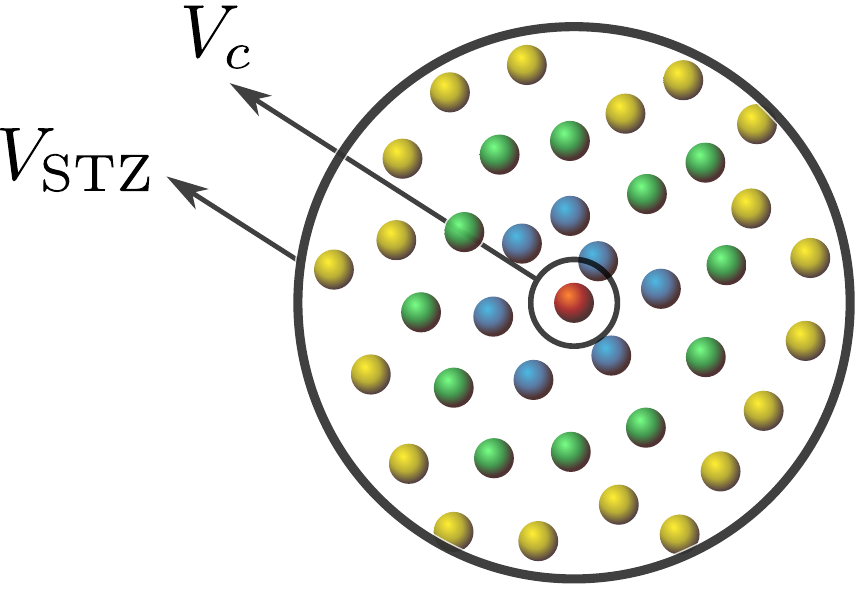}
	\caption{Illustration of a cut through an idealized spherical STZ. The diameter of the STZ is roughly six times larger than the diameter of the characteristic atomic volume $V_c$.  The coloured spheres represent the average ionic cores of the atoms in the alloy. The objects are approximately drawn to scale.\label{STZsketch}}
\end{figure}
\subsection{Cooperative shear events}
The constitutive equation for the viscosity presented in the main text is based on the shoving model of glasses~\cite{Dyre1998, Dyre1996}. The key idea behind this is that the interatomic repulsion is strong compared to the attractive forces. In a plastic flow event many atoms need to rearrange, which at constant volume has the consequence that the atoms involved are forced together~\cite{Dyre1996}. Since this process is energetically unfavourable, it was suggested that the atoms may shove aside their neighbours to create free volume and accommodate the flow process.

It can be shown that the energy necessary to trigger a shoving event in a volume $V$ is linearly proportional to the high-frequency shear modulus, with the  constant of proportionality being the characteristic volume $V_c$, i.e. $
E(T) = G_{\text{A}}(T) V_c
$~\cite{Dyre1996}.

According to Ref.~\cite{Johnson2005}, the total energy barrier $W$ between two basins in the potential energy landscape is given by
$
W=\frac{8}{\pi^2}
G_{\text{A}}
\, \Omega
\gamma_C^2 \xi,
$
where $\Omega$ represents the effective STZ volume and $G_{\text{A}}$ the shear modulus.
The critical shear strain $\gamma_C$ is found be $0.036\pm 0.002$ for all known metallic glasses and the correction factor $\xi$ due to the matrix confinement of a STZ ranges between 2 and 4~\cite{Johnson2005}.

After evaluating the expression for the potential energy barrier we get $W \approx (1/320) \,G_{\text{A}} \Omega$. 

We assume that the energy barrier $W$ in the cooperative shear model is approximately equal to the shoving energy, that is $W \approx E$. It directly follows that $G_{\text{A}} V_c \approx (1/320) G_{\text{A}} \Omega$, which leads us to conclude that the effective volume of a STZ is  $\Omega \approx 320 V_c$. It is physically meaningful that the effective STZ volume $\Omega$ is approximately equivalent to the initial shoving volume $V$, henceforth calling it $V_{\text{STZ}}$.
Using the identification $\Omega \approx V_{\text{STZ}}\approx320 V_c$ together with Eq.~(8) from the main text, we obtain a relation  between the activation volume and the characteristic volume $V_c$ given by $\Delta V \approx \sqrt{480} \,V_c$. 

As explained in the main text, we use the characteristic volumes $V_c$ from the viscosity fitting in the relation above to compare the resulting theoretical values for the STZ volume to experimental data.  The relative sizes of the STZ and the characteristic volume $V_c$ are sketched for a spherical geometry in Fig.~\ref{STZsketch}.

\begin{table}[b]\label{tab2}
	\caption{Values used for the fitting of the Ashcroft-Born- Mayer pseudopotential to the logarithmic potential of mean force \label{potentialfitting}}
	\begin{tabular}{@{ \vrule height 10.5pt depth4pt  width0pt}lccccc}
		\hline
		$\lambda$ &$A$ (eV) &$q_{\text{TF}}$ $(\mathring{A}^{-1})$&$B$ (eV)&$C$ $(\mathring{A}^{-1})$ &$\bar{\sigma}$ $(\mathring{A})$
		\\
		\hline
		196.2
		&0.37	&1.7	&5.1	&2.8& 1.68\\
		276.4
		&0.39	&1.7&21.8	&2.8& 1.29\\
		
		286.5
		&0.42	&1.7	&28.8	&2.8&1.20\\
		381.2
		&0.48	&1.7	&57.4	&2.8&1.056\\ 
		\hline
	\end{tabular}
\end{table}
\end{article}

\end{document}